\documentclass[11pt]{article}
\usepackage{a4}
\usepackage{amsmath}
\usepackage{amssymb}
\usepackage[dvips]{graphicx}
\usepackage{subfigure}
\def\beq{\begin{equation}}
\def\eeq{\end{equation}}
\def\beqa{\begin{eqnarray}}
\def\eeqa{\end{eqnarray}}
\def\n{\nonumber \\}

\newcommand {\tr}{{\rm tr}\,}

\newcommand {\CTr}{{\cal T}r\,}

\newcommand {\del}{\partial}

\newcommand {\al}{\alpha}
\newcommand {\be}{\beta}

\begin{document}

\begin{flushright}
{SAGA-HE-256}\\
{KEK-TH-1331}
\end{flushright}
\vskip 0.5 truecm

\begin{center}
{\large{\bf 
Construction of a topological charge on 
fuzzy $S^2 \times S^2$ \\
via Ginsparg-Wilson relation
}}
\vskip 1.0cm

{\large Hajime Aoki$^{a}$\footnote{e-mail
 address: haoki@cc.saga-u.ac.jp},
 Yoshiko Hirayama$^{b}$\footnote{e-mail
 address: hirayama@tenor.ocn.ne.jp}
 and
 Satoshi Iso$^{c}$\footnote{e-mail
 address: satoshi.iso@kek.jp}
}
\vskip 0.5cm

$^a${\it Department of Physics, Saga University, Saga 840-8502,
Japan  }\\
$^b${\it Miyazaki Information Processing Center Limited, 
Fukuoka 812-0011, Japan}\\
$^c${\it KEK Theory Center, 
High Energy Accelerator Research Organization (KEK)  \\
and 
the Graduate University for Advanced Studies (SOKENDAI), \\
Ibaraki 305-0801, Japan }
\end{center}

\vskip 1cm
\begin{center}
\begin{bf}
Abstract
\end{bf}
\end{center}
We construct a topological charge of gauge field
configurations on a
fuzzy $S^2 \times S^2$ by using a Dirac operator 
satisfying the Ginsparg-Wilson relation. 
The topological charge defined on the fuzzy $S^2 \times S^2$
can be interpreted as a noncommutative (or matrix) 
generalization of
the 2nd Chern character on $S^2 \times S^2$.
We further calculate the number of chiral zero modes of the Dirac
operator in  topologically nontrivial gauge configurations.
Generalizations of our formulation to fuzzy 
$(S^2)^k$ are also discussed.

\newpage
\section{Introduction}   
\setcounter{footnote}{0}
\setcounter{equation}{0}

Noncommutative geometry \cite{Connes} 
appears naturally in string theory \cite{CDS,NCMM,Seiberg:1999vs},
and is also encoded 
in the matrix model formulations of 
the string theory \cite{Banks:1996vh,IKKT}.
In the superstring theory,
the size of six dimensions is expected to become tiny
and the ten-dimensional spacetime becomes compactified to
four dimensions. 
Then the number of massless fermions, 
in particular, the number of generations 
in the four-dimensional spacetime is 
given by the topology of the six-dimensional compactified space. 
Then, if the size of the compactified space is as small as
the Planck scale, its coordinates may become
noncommutative and we will need
to generalize the notion of topology to
noncommutative spaces.

In ordinary spaces, 
the topological charge of gauge field configurations 
can be  provided by the index of 
the Dirac operator, {\it i.e.}, 
the difference of the numbers of chiral zero modes,
via the index theorem \cite{Atiyah:1971rm}.
Generalizations of the index theorem 
to noncommutative spaces are,
however, mostly formulated
in spaces with an infinite size,
and it is widely believed that topological charges
cannot be defined in a system with  finite degrees of freedom.

The situation is similar to the lattice gauge theories,
where the theory is defined on a finite number of 
lattice points and the total degrees of freedom are finite.
There a problem to properly define the chiral symmetry and 
the index theorem arises
due to the doubling problem
\cite{Nielsen:1980rz}.
The problem has been solved successfully 
by introducing Dirac operators
satisfying a Ginsparg-Wilson (GW)
relation \cite{GinspargWilson}.
While all the gauge field configurations are continuously 
connected and there seems to be no room for defining 
separate topological sectors in such systems 
with finite degrees of freedom,
the configuration space becomes disconnected by 
introducing the admissibility condition, 
and the various topological sectors
can then be realized
\cite{Luscher:1981zq}.

In a previous paper \cite{AIN2}, we have proposed to use 
the GW relation to define a topological charge and 
to classify the gauge field configurations in 
noncommutative spaces with finite degrees of freedom.
We have provided 
a general prescription to construct 
a GW Dirac operator with a coupling to background gauge fields.
As a concrete example,
a GW Dirac operator on the fuzzy $S^2$
was given.
(See also \cite{balagovi}
for an earlier construction of the GW Dirac operator
on fuzzy $S^2$ without the background gauge field.)\footnote{
In the case of noncommutative tori, the gauge fields are
represented by unitary matrices of Wilson lines and 
a GW Dirac operator can be constructed similarly to 
the lattice gauge theory.
It was given in \cite{Nishimura:2001dq} 
and analyzed in \cite{Iso:2002jc}.
For constructions of the GW Dirac operators 
in gauge field backgrounds 
with nontrivial topology,
see \cite{AIM,Aoki:2008qta} for fuzzy $S^2$ and
\cite{Aoki:2008ik} for noncommutative tori.}

In this paper, 
we further apply the proposal in ref.[11] to fuzzy 
$S^2 \times S^2$.
We first construct a GW Dirac operator
on fuzzy $S^2\times S^2$.\footnote{
A Dirac operator on fuzzy $S^2\times S^2$
without the GW relation
was given in \cite{Behr:2005wp}.
Dynamics of gauge theory on fuzzy $S^2\times S^2$
was studied in \cite{Imai:2003ja}.}
Owing to the GW relation, 
the topological charge is given by the 
index of the Dirac operator.
We then study the commutative limit of the
topological charge. It becomes
a sum of the 2nd Chern character on $S^2\times S^2$
and the 1st Chern character.
We also investigate the chiral zero modes of the Dirac operator
for some specific gauge field backgrounds and confirm that
the index of the Dirac operator takes the consistent values.
We finally generalize our formulation to fuzzy $(S^2)^k$.

The paper is organized as follows.
After briefly reviewing the GW relation on fuzzy $S^2$
in section \ref{sec:review_S2}, 
we construct a GW Dirac operator on fuzzy $S^2\times S^2$
in section \ref{sec:Form-S2S2}.
In section \ref{sec:com_lim_top_char}, 
we calculate the commutative limit of the topological charge.
We then study the chiral zero modes of the Dirac operator
for the free case in section \ref{sec:zero-mode_free},
and for the monopole backgrounds in section \ref{sec:monopole}.
Here we also introduce a projected topological charge that 
gives correct values for topologically nontrivial gauge field configurations.
Generalizations of our formulation to fuzzy $(S^2)^k$
are given in section \ref{sec:higherdimension}.
Section \ref{sec:conclusion} is devoted 
to conclusions and discussions.
In appendices
\ref{sec:calOfTrG} and 
\ref{sec:cal2ndChern}, we give detailed calculations 
of the commutative limit.
In appendix \ref{sec:Spectrum_Free}, 
a full spectrum of the Dirac operator for the free case
is obtained.
A calculation of the topological charge
for a modified Dirac operator is given in 
appendix \ref {sec:one_time_normalization}.

\section{Brief review of GW relation on fuzzy $S^2$}
\label{sec:review_S2}
\setcounter{equation}{0}
We first briefly review the Ginsparg-Wilson (GW) relation 
on a fuzzy $S^2$, following the prescription given in
ref.\cite{AIN2}.

Noncommutative coordinates of 
fuzzy $S^2$ are given by
$x_i =\mu L_i$, 
where $\mu$ is a noncommutative parameter,
and $L_i$ is the $n$-dimensional irreducible
representation matrix of the $SU(2)$ algebra.
Then we have the relation
$
(x_i)^2
=\mu^2 \frac{n^2-1}{4}\mbox{\boldmath  $1$}_n 
= \rho^2 \mbox{\boldmath  $1$}_n 
$,
where 
$\rho=\mu \sqrt{(n^2-1)/4}$ 
expresses the radius of the $S^2$.
The commutative limit is taken by 
$\mu \to0, n \to \infty$
with $\rho$ fixed.

In our formulation of the GW relation,
we first define two chirality operators as
\beqa
\Gamma_X &=& a\left(\sigma_i L_i^R -\frac{1}{2}\right)_X \ , 
\label{def_GforS2}  \\
\hat\Gamma_X &=& \frac{H_X}{\sqrt{H^2_X}}  \ , \hspace{0.5cm}
H_X=a\left(\sigma_i A_i +\frac{1}{2}\right)_X  \ ,
\label{def_hatGforS2} 
\eeqa
with covariant coordinates
\beq
(A_{i})_X = (L_{i}+\rho a_{i})_X \ .
\label{defcovdel}
\eeq
The subscript $X=1,2$ will be used for labeling each $S^2$ 
of $S^2 \times S^2$ in the following sections,
and it can be ignored in the present section.
The superscript $R$ in $L_i^R$ means that this 
operator acts from the right on matrices, 
while the other operators without the superscript $R$
act from the left.
The number $a=2/n$ serves as a noncommutative analog
of the lattice spacing,
and $\sigma_i$ is the Pauli matrix.
The matrices $a_i$ in (\ref{defcovdel}) represent 
the gauge field, and the gauge transformation for the
covariant coordinate is given by $A_i \rightarrow UA_i U^\dagger$.
The fermionic fields $\psi$ on which these chiral operators act
are in the fundamental representation of the gauge group,
and the gauge transformation is given by $\psi \rightarrow U\psi.$
Hence, both $\Gamma_X \psi$ and $\hat{\Gamma}_X \psi$ 
transform covariantly 
as $\Gamma_X \psi \rightarrow U \Gamma_X \psi$
and $\hat\Gamma_X \psi \rightarrow U \hat\Gamma_X \psi$.
$U(n)$ gauge symmetry can be realized 
 by taking
$L_i = L_i \otimes {\bf 1}$
and
$a_i = a_i^a T^a$,
where $T^a$'s are the generators of $U(n)$
and $a_i^a$'s are functions of the coordinates $L_i$.

From the definitions 
(\ref{def_GforS2}) and (\ref{def_hatGforS2}),
the chirality operators 
satisfy the relations
\beq
(\Gamma_X)^\dagger=\Gamma_X \ , \ \ \
(\hat\Gamma_X)^\dagger=\hat\Gamma_X \ , \ \ \
(\Gamma_X)^2=(\hat\Gamma_X)^2=1 \ .
\label{G2_HG2_Gd_HGd_X}
\eeq
One can also show that
in the commutative limit, both $\Gamma_X$ and $\hat\Gamma_X$
become the same chirality operator $\gamma_X = (n_i \sigma_i)_X$
on a commutative $S^2$
where
$(n_i)_X=(x_i)_X/\rho$ is a unit vector on $S^2$.

We next define a GW Dirac operator by
\beq
(D_{\rm GW})_X = -a^{-1} (\Gamma- \hat{\Gamma})_X \ .
\label{defDGWX}
\eeq
It satisfies the GW relation  
\beq
(\Gamma D_{\rm GW}+D_{\rm GW} \hat{\Gamma})_X=0 \ .
\label{eqn:GWrelationX}
\eeq
Hence, the index, {\it i.e.}, the difference of the numbers 
of the chiral zero modes, is given by
the trace of the chirality operators as
\beq
{\rm index}((D_{\rm GW})_X) = 
\frac{1}{2} \CTr [\Gamma + \hat\Gamma]_X \ .
\label{eqn:indextheogenX}
\eeq
Here $\CTr$ is the trace in the whole configuration space,
that is, over the spinorial index, the gauge group index,
and the matrix space representing the coordinates.
Since the definition of $\hat\Gamma_X$ depends on the gauge
field backgrounds, 
the right-hand side (rhs)
of (\ref{eqn:indextheogenX}) gives a noncommutative
generalization of the topological charge.
Thus, eq.(\ref{eqn:indextheogenX}) gives
an index theorem on fuzzy $S^2$.

In the commutative limit,
the Dirac operator (\ref{defDGWX}) becomes
\beq
(D_{\rm GW})_X \to
D'_X = \Bigl(\sigma_i ({\cal L}_i + \rho P_{ij} a_j ) +1\Bigr)_X \ ,
\label{comlim_DGW_X}
\eeq
where ${\cal L}_i=-i\epsilon_{ijk} x_j \partial_k$'s 
are the derivative operators along the Killing vectors on $S^2$,
and $P_{ij}=\delta_{ij}-n_i n_j$ 
is the projection operator on the tangential directions 
on $S^2$.
The tangential components of the gauge field $a_i$ represent
the gauge field on $S^2$ while the normal component
becomes a scalar field $\phi = n_i a_i$.
Because of the GW relation, the Dirac operator
is not coupled to the scalar field,
since such a coupling would violate the chiral symmetry
on $S^2$ and contradict with the GW relation.

The commutative limit of the topological charge,
the rhs of (\ref{eqn:indextheogenX}), is shown to become
\cite{AIN2,AIN3}
\beq
\frac{1}{2} \CTr [\Gamma + \hat\Gamma]_X
\to
\rho^2 
\left( \int \frac{d\Omega}{4\pi} 
\tr (\epsilon_{ijk} n_{k}F_{ij}) \right)_X \ ,
\label{eqn:1stChernX}
\eeq
where $\tr$ is the trace over the gauge group.
The field strength $F_{ij}$ is defined as
$F_{ij}= \partial_i a_j'-\partial_j a_i'-i[a_i',a_j']$,
where $a'_i$ is the tangential components of the gauge field,
given as
$a'_i = \epsilon_{ijk}n_j a_k$.
This is the integral of 
the 1st Chern character on a commutative $S^2$.

In order to construct  topologically nontrivial configurations,
we need a bit more modification
\cite{Balachandran:2003ay, AIN3, AIM, Aoki:2008qta}.  
Consider, for instance, $U(2)$ gauge theory on the fuzzy $S^2$.
Then some gauge field configurations  $a_i$ 
break the $U(2)$ gauge symmetry to $U(1) \times U(1)$. 
They correspond to  nontrivial elements of
$\Pi_2(SU(2)/U(1))$ and 
physically to the 't Hooft-Polyakov-type monopoles. 
A topological charge can be also constructed
by modifying the index theorem
(by inserting a projection operator), and 
it correctly reproduces the topological
charge of such configurations. 
This issue is discussed later
in section \ref{sec:monopole} 
for the case of fuzzy $S^2 \times S^2$.

\section{GW relation on fuzzy $S^2 \times S^2$}
\label{sec:Form-S2S2}
\setcounter{equation}{0}

We now construct a GW Dirac operator and the 
corresponding topological charge 
on  fuzzy $S^2 \times S^2$.

As in fuzzy $S^2$,
we first define two chirality operators as  
\beqa
\Gamma &=& \Gamma_1 \Gamma_2 \ , 
\label{eqn:def_G} \\
\hat\Gamma &=&
\frac{\{ \hat\Gamma_1, \ \hat\Gamma_2 \}}
{\sqrt{\{ \hat\Gamma_1, \ \hat\Gamma_2 \}^2}} \ ,
\label{eqn:def_HatG}
\eeqa
where $\Gamma_X$ and $\hat\Gamma_X$ with $X=1,2$
are the chirality operators on each fuzzy $S^2$
labeled by $X$.
They are given in 
(\ref{def_GforS2}) and (\ref{def_hatGforS2}).
For simplicity, we take the radii of the two spheres equal.
Note that while the index $i$ of the gauge field $(a_i)_X$ 
refers to each $S^2$ labeled by $X$,
the gauge field depends 
on the coordinates of both $S^2$'s, $(L_i)_1$ and $(L_i)_2$.

From (\ref{def_GforS2}) and (\ref{def_hatGforS2}),
one has
\beq
[\Gamma_1, \Gamma_2] = [\Gamma_1, \hat\Gamma_2]
=[\hat\Gamma_1, \Gamma_2]=0 \ .
\label{gamma12commute}
\eeq
One can also show from (\ref{G2_HG2_Gd_HGd_X}) that
\beq
\{ \hat\Gamma_1, \ \hat\Gamma_2 \}^2
= 4+ [ \hat\Gamma_1, \ \hat\Gamma_2 ]^2 \ ,
\label{relanticomcom}
\eeq
where the second term is of order ${\cal O}(n^{-4})$,
as is shown below (\ref{comhatgam12}).

From the relation of the chirality operator 
(\ref{G2_HG2_Gd_HGd_X}) on each 
sphere,
the chirality operators 
(\ref{eqn:def_G}) and (\ref{eqn:def_HatG}) on $S^2 \times S^2$ also
satisfy the same relations
\beq
(\Gamma)^\dagger=\Gamma \ , \ \ \
(\hat\Gamma)^\dagger=\hat\Gamma \ , \ \ \
(\Gamma)^2=(\hat\Gamma)^2=1 \ .
\label{eqn:Gh_G^2_relation}
\eeq
One can also show that
in the commutative limit, both operators, $\Gamma$ and $\hat\Gamma$,
become the same chirality operator $\gamma =\gamma_1 \gamma_2$
on a commutative $S^2 \times S^2$.
The second term of (\ref{relanticomcom}) does not 
contribute to the commutative limit of (\ref{eqn:def_HatG})
because of the ${\cal O}(n^{-4})$ behavior.
It should be, however, noted that this term is relevant in calculating 
the commutative limit of the topological charge.

We then define a GW Dirac operator as
\beq
D_{\rm GW} = -a^{-1} (\Gamma- \hat{\Gamma}) \ ,
\label{defDGW}
\eeq
which satisfies the GW relation  
\beq
\Gamma D_{\rm GW}+D_{\rm GW} \hat{\Gamma}=0 
\label{eqn:GWrelation}
\eeq
and the index theorem 
\beq
{\rm index}(D_{\rm GW}) = 
\frac{1}{2} {\cal T}r [\Gamma + \hat\Gamma] \ ,
\label{eqn:indextheogen}
\eeq
where
$\CTr$ is the trace over  the whole configuration space,
that is, over the spinorial indices of both spheres,
the gauge group index,
and the matrix space spanned by 
polynomials of the coordinates $(L_i)_1$ and  $(L_i)_2$.

The commutative limit of the Dirac operator can be similarly
obtained.
Using the relation
\beq
\Gamma_1\Gamma_2 - \hat\Gamma_1\hat\Gamma_2
=\frac{1}{2}\Bigl[(\Gamma_1 -\hat\Gamma_1)(\Gamma_2 +\hat\Gamma_2)
+(\Gamma_1 +\hat\Gamma_1)(\Gamma_2 -\hat\Gamma_2) \Bigr] \ ,
\label{eqn:G-Grel}
\eeq
and (\ref{comlim_DGW_X}),
one can show that in the commutative limit
the GW Dirac operator (\ref{defDGW}) becomes 
\beq
D_{\rm GW} \to
D'_{1} \gamma_2+\gamma_1 D'_{2} \ ,
\label{eqn:DGWcom}
\eeq
where $D'_X$ and $\gamma_X$ are Dirac and chirality operators
on each $S^2$. 
This is not exactly the same as the ordinary 
Dirac operator on 
a commutative $S^2 \times S^2$
\footnote{
Taking the planar limit at the north pole
$(n_i)_{X=1}=(n_i)_{X=2}=\delta_{i,3}$,
the 4 dimensional gamma matrices become
$\gamma_1=(\sigma_1)_{X=1} (\sigma_3)_{X=2}, \ 
\gamma_2=(\sigma_2)_{X=1} (\sigma_3)_{X=2}, \ 
\gamma_3=(\sigma_3)_{X=1}(\sigma_1)_{X=2}, \ 
\gamma_4=(\sigma_3)_{X=1}(\sigma_2)_{X=2}$, 
and they  do not satisfy the $SO(4)$ Clifford algebra. 
However, 
if one multiplies the GW Dirac operator (\ref{defDGW})
by $\Gamma_1$ from the left in the definition,
for instance,
then in the commutative limit, 
the gamma matrices are multiplied by 
$(\sigma_3)_{X=1}$ from the left, 
giving
$\tilde\gamma_1=i(\sigma_2)_{X=1} (\sigma_3)_{X=2}, \ 
\tilde\gamma_2= -i(\sigma_1)_{X=1} (\sigma_3)_{X=2}, \ 
\tilde\gamma_3= (\sigma_1)_{X=2}, \ 
\tilde\gamma_4= (\sigma_2)_{X=2}$, 
which satisfy $SO(2,2)$ Clifford algebra.
},
but we will show later that 
the Dirac operator (\ref{defDGW}) suffices to
define a topological charge on fuzzy $S^2 \times S^2$.

Our formulation has the following nice
properties.
First, it is manifestly covariant under the
gauge transformation
\beq
(A_i)_X \to U \, (A_i)_X \, U^\dagger 
\eeq
for both $X=1,2$ with a common $U$,
which is a general unitary matrix depending on 
the coordinates of both spheres, $(L_i)_1$ and $(L_i)_2$.
Second, the GW relation assures the
topological property of the index and the topological charge.
Finally, 
the formulation 
has manifest $SO(3) \times SO(3)$ Poincare 
invariance on $S^2 \times S^2$.
Because of these properties, the commutative limit
of the topological charge we have
defined should become a sum of the 1st and the 2nd Chern characters
on $S^2 \times S^2$. 
This is what we will show in the next section.

\section{Commutative limit of the topological charge}
\label{sec:com_lim_top_char}
\setcounter{equation}{0}

In this section, 
we calculate the commutative limit of the topological charge
defined in the rhs of (\ref{eqn:indextheogen}).
As we discussed at the end of the previous section,
the result should be a linear combination
of a constant, the 1st Chern character and the 2nd Chern character.

$\CTr [ \Gamma ] $ is easily calculated as
\beq
\CTr [ \Gamma ] = 4n^{2} \tr({\bf 1}) \ ,
\label{eqn:TrG} 
\eeq
where $\tr$ is the trace over the gauge group space.

On the contrary, the evaluation of $\CTr[\hat\Gamma]$ is
more involved.
As we show in Appendix \ref{sec:calOfTrG}, 
by expanding it in the gauge fields,
it becomes a sum of five terms if we take terms up to order $n^{-4}$:
\beq
\CTr [ \hat\Gamma] 
= \CTr \left[ \sum_{i=1}^5 G_i+{\cal O}(n^{-5})
\right]  
\label{hatG_exp_n-1_-4}.
\eeq
The terms of order ${\cal O}(n^{-5})$ vanish in the commutative limit,
since the trace $\CTr$ gives  a contribution of order $n^4$.
Each term is given by
\beqa
G_1 &=& \al_1 \al_2  \ ,
\label{eqn:TrHatG_0}\\
G_2 &=& \frac{1}{2} \left(
\{\al_1, \ \zeta^{(1)}_2+\zeta^{(2)}_2 \} 
+\{\al_2, \ \zeta^{(1)}_1+\zeta^{(2)}_1 \} \right) \ , 
\label{eqn:TrHatG_1_1stChern1}\\
G_3 &=& \frac{1}{2} \left(
\{\al_1, \ \zeta^{(3)}_2 \} 
+\{\al_2, \ \zeta^{(3)}_1 \} \right) \ , 
\label{eqn:TrHatG_2_1stChern2_1}  \\
G_4 &=&  \frac{1}{2} \left(
\{ \zeta^{(1)}_1+\zeta^{(2)}_1, \ \zeta^{(1)}_2+\zeta^{(2)}_2 \} 
\right) \ , 
\label{eqn:TrHatG_2_2ndChern1}\\
G_5 &=&  -\frac{1}{8}\al_1 \al_2
\left([\al_1, \ \zeta^{(1)}_2 ]-[\al_2, \ \zeta^{(1)}_1 ]
+[\zeta^{(1)}_1, \ \zeta^{(1)}_2 ]\right)^2 \ ,
\label{eqn:TrHatG_2_2ndChern2}  
\eeqa
where $\al_X$ and $\zeta^{(i)}_X$ 
are $0$-th and $i$-th order in the gauge field $(a_i)_X$,
and are defined by
(\ref{eqn:def_al_be}) and
(\ref{hatGX_exp_a1})-(\ref{hatGX_exp_a3}).
The last term $G_5$
comes from the denominator of 
(\ref{eqn:def_HatG}).
Contrary to the commutative limit of the chirality operators
or the Dirac operator, we should take care of the  
order ${\cal O}(n^{-4})$ term from the
denominator.

The first term $\CTr [G_1]$ becomes a constant
\beq
\CTr [G_1] = 4n^{2} \tr({\bf 1}) \ .
\eeq
It is the same as (\ref{eqn:TrG}).
The commutative limit of 
$\CTr [G_2]$
can be calculated as in (\ref{eqn:1stChernX})
for the fuzzy $S^2$, and gives terms  proportional
to the 1st Chern character on each sphere:
\beq
\CTr [G_2]
\to
2n \cdot 2\rho^2 
\int \frac{d\Omega_1}{4\pi} \frac{d\Omega_2}{4\pi}
\tr(\epsilon_{abc} n_{c}F_{ab}+\epsilon_{ijk} n_{k}F_{ij}) \ .
\label{eqn:1stChern}
\eeq
The indices $a$, $b$, and $c$ refer to the first $S^2$,
while the indices $i$, $j$, and $k$  refer to the second $S^2$.
Note, however, that the field strength, 
$F_{ab}(\Omega_1$, $\Omega_2)$ and 
$F_{ij}(\Omega_1$, $\Omega_2)$,
can depend on the coordinates of both $S^2$.
In this sense, 
(\ref{eqn:1stChern}) represents a generalized
1st Chern character 
defined on a commutative $S^2\times S^2$.  
Since (\ref{eqn:1stChern}) is of order $n$,
the subleading order terms in $n^{-1}$ in $G_2$
give a finite contribution.
The commutative limit of $\CTr [G_3]$ also gives
a finite contribution.
Since these terms vanish for the configurations
that will be discussed later, we do not write these terms explicitly
in this paper.
We will study topological charges for 
more general configurations in a separate paper.

The commutative limit of $\CTr [G_4]$
can also be calculated as in (\ref{eqn:1stChernX})
and becomes
\beq
\CTr [G_4] \to
(2\rho^2)^2 
\int \frac{d\Omega_1}{4\pi} \frac{d\Omega_2}{4\pi}
\tr(\epsilon_{abc} n_{c}F_{ab}\epsilon_{ijk} n_{k}F_{ij}) \ .
\label{eqn:2ndChern_1}
\eeq
Remarkably, as we show in Appendix \ref{sec:cal2ndChern}, 
the commutative limit of $\CTr [G_5]$
becomes 
\beq
\CTr [G_5]
\to
-8\rho^4
\int \frac{d\Omega_1}{4\pi} \frac{d\Omega_2}{4\pi}
\tr(\epsilon_{abc} n_{c}\epsilon_{ijk} n_{k}F_{ai}F_{bj}) \ .
\label{eqn:2ndChern_2}
\eeq
Note that the field strengths with indices from different
spheres, $F_{ai}$ and $F_{bj}$, arise here. 
Combining these two terms we obtain
\beq
\CTr[G_4 + G_5]
\to
4\rho^4 
\int \frac{d\Omega_1}{4\pi} \frac{d\Omega_2}{4\pi}
\epsilon_{abc} n_{c}\epsilon_{ijk} n_{k}
\tr(F_{ab}F_{ij}-F_{ai}F_{bj}+F_{aj}F_{bi}) \ .
\label{eqn:2ndChern_all}
\eeq
This gives an integral of the 2nd Chern character 
on a commutative $S^2 \times S^2$.

To summarize,
the commutative limit of the topological charge on $S^2 \times S^2$ 
becomes
\beqa
&&\frac{1}{2}\CTr[\Gamma + \hat\Gamma] \n
&\to&
4n^2 \tr({\bf 1})
+2n\rho^2 
\int \frac{d\Omega_1}{4\pi} \frac{d\Omega_2}{4\pi}
\tr(\epsilon_{abc} n_{c}F_{ab}+\epsilon_{ijk} n_{k}F_{ij}) \nonumber \\
&&+2\rho^4 
\int \frac{d\Omega_1}{4\pi} \frac{d\Omega_2}{4\pi}
\epsilon_{abc} n_{c}\epsilon_{ijk} n_{k}
\tr(F_{ab}F_{ij}-F_{ai}F_{bj}+F_{aj}F_{bi}) \ . 
\label{ComLimTopCharResult}
\eeqa
In the differential forms,  
it is rewritten as
\beq
\frac{1}{(2\pi)^2} \int \tr\left(
n^2 d\Omega_1 d\Omega_2 +n(d\Omega_1 F_2 +d\Omega_2 F_1)  + 
2\frac{1}{2!}(F^2)_{12}\right) \ ,
\label{ComLimTopCharResult_form}
\eeq
with
\beqa
F_X &=& \frac{1}{2}\rho^2 (d\Omega \ 
\epsilon_{ijk} n_{k} F_{ij})_X \ , 
\label{defFX}\\
(F^2)_{XY} &=& \frac{2!}{2^2} \rho^4 d\Omega_X d\Omega_Y
\Bigl(\epsilon_{abc} n_{c}\epsilon_{ijk} n_{k}
(F_{ab}F_{ij}-F_{ai}F_{bj}+F_{aj}F_{bi})\Bigr)_{XY} \ .
\label{defF2XY}
\eeqa
Here $d\Omega_X$ is the volume form on each $S^2$.
In the flat limit, $F$ and $F^2$ become  familiar forms
on each $S^2$ and $S^2 \times S^2$ respectively:
\beqa
F_X &\to& 
\frac{1}{2} (F_{\mu\nu}dx_\mu \wedge dx_\nu )_X \ , \\
(F^2)_{XY} &\to& 
\frac{1}{2^2} (F_{\mu\nu} F_{\lambda\rho}
dx_\mu \wedge dx_\nu \wedge dx_\lambda \wedge dx_\rho)_{XY} \ .
\eeqa

The first and the second terms 
in (\ref{ComLimTopCharResult})
and (\ref{ComLimTopCharResult_form})
are proportional to 
$n^2$ and $n$ respectively, 
and they diverge in the commutative (large $n$) limit.
The third term is also twice the 2nd Chern character,
and the topological charge we have defined by the 
GW Dirac operator is different from the 
index of the ordinary Dirac operator on $S^2 \times S^2$.
This is because the Dirac operator is different from 
the ordinary one as we discussed 
below (\ref{eqn:DGWcom}).
We will discuss origins
of each term in (\ref{ComLimTopCharResult})
and (\ref{ComLimTopCharResult_form})
by investigating the chiral zero modes
in the following sections.

While the topological charge defined this way
contains various topological invariants,
we can, nevertheless, extract the 2nd Chern character.
In order to define a noncommutative analog 
of the 2nd Chern character on $S^2 \times S^2$, 
we  subtract the extra pieces as follows:
\beq
\frac{1}{4}\CTr[\Gamma + \hat\Gamma]
-\frac{1}{4n}\CTr[\Gamma_1 + \hat\Gamma_1]
-\frac{1}{4n}\CTr[\Gamma_2 + \hat\Gamma_2]
-\frac{1}{2n^2}\CTr[{\bf 1}] \ .
\label{def_top_charge_subtract}
\eeq
Each term is a topological invariant on fuzzy $S^2 \times S^2$
and is well-defined on it.


\section{Chiral zero modes}
\label{sec:chiral_zero}

In this section, we explicitly calculate the number
of chiral zero modes in some specific configurations
and compare it with the topological charge in the commutative limit,
(\ref{ComLimTopCharResult})
or (\ref{ComLimTopCharResult_form}). 
Especially we discuss why the index explicitly
depends on the size $n$ of the matrices.

\subsection{Chiral zero modes for the free case}
\label{sec:zero-mode_free}
\setcounter{equation}{0}

We first investigate the chiral zero modes of
the GW Dirac operator 
for the free case where the gauge field vanishes.
Even in the absence of the gauge field, 
there exist chiral zero modes of the GW Dirac operator
and they give the first term of 
(\ref{ComLimTopCharResult})
or (\ref{ComLimTopCharResult_form}).
We here consider $U(1)$ gauge group, for simplicity.

In the free case, 
we have a simple relation $[\hat\Gamma_1, \hat\Gamma_2]=0$,
and the chirality operator (\ref{eqn:def_HatG}) can be simplified
as $\hat\Gamma = \hat\Gamma_1 \hat\Gamma_2$.
Using  
the relation (\ref{eqn:G-Grel}),
the GW Dirac operator (\ref{defDGW}) is also simplified as
\beq
D_{\rm GW} = D_1 +D_2
\label{eqn:DGW}
\eeq
where
\beq
D_1=-\frac{1}{2}a^{-1}(\Gamma_1 -\hat\Gamma_1)
(\Gamma_2 +\hat\Gamma_2) \ , \ \
D_2=-\frac{1}{2}a^{-1}(\Gamma_1 +\hat\Gamma_1)
(\Gamma_2 -\hat\Gamma_2) \ .
\label{def_D1D2_free}
\eeq
Using (\ref{gamma12commute}), $[\hat\Gamma_1, \hat\Gamma_2]=0$,
and (\ref{G2_HG2_Gd_HGd_X}),
one can easily show the following GW relations for each $D_a$: 
\beq
\Gamma D_1 + D_1 \hat\Gamma =0 \ , \ \
\Gamma D_2 + D_2 \hat\Gamma =0 \ ,
\label{GWrelD1D2}
\eeq
where $\Gamma$ and $\hat\Gamma$ are
the chirality operators on the fuzzy $S^2\times S^2$
defined in (\ref{eqn:def_G}) and (\ref{eqn:def_HatG}).
One can also show 
\beq
[D_1 , \  D_2]=0 \ . 
\label{D1D2commute}
\eeq

Now consider states with zero eigenvalues of 
the Dirac operator $D_{GW}$.
The chirality operators can also be diagonalized in this space
owing to the GW relation (\ref{eqn:GWrelation}).
Hence we  consider a state  
$|\psi \rangle$ satisfying
\beq
D_{\rm GW} |\psi \rangle =0 \ , \ \
\Gamma |\psi \rangle = \hat\Gamma |\psi \rangle 
= \pm |\psi \rangle \ .
\eeq
Then from 
(\ref{GWrelD1D2}) and (\ref{D1D2commute}), we have
\beq
D_{\rm GW} D_a|\psi \rangle =0 \ , \ \
\Gamma D_a|\psi \rangle = \hat\Gamma D_a|\psi \rangle 
= \mp D_a|\psi \rangle \ ,
\eeq
for $a=1,2$.
Therefore, if either
$D_1 |\psi \rangle \neq 0$ or $D_2 |\psi \rangle \neq 0$
is satisfied,
the contributions to the index of $D_{\rm GW}$ cancel
each other by $|\psi \rangle$ and $D_a |\psi \rangle$.
Thus
a chiral zero mode that can contribute to the index 
must satisfy
$D_1 |\psi \rangle = 0$
and $D_2 |\psi \rangle = 0$.
From (\ref{def_D1D2_free}),
a zero mode of $D_1$ is given by 
a zero mode of $\Gamma_1 -\hat\Gamma_1$
or a zero mode of $\Gamma_2 +\hat\Gamma_2$,
owing to $[\Gamma_1 -\hat\Gamma_1, \ \Gamma_2 +\hat\Gamma_2]=0$.
Similarly, a zero mode of $D_2$ is given by
that of $\Gamma_1 + \hat\Gamma_1$ or 
 $\Gamma_2 -\hat\Gamma_2$.

We then study each fuzzy $S^2$ separately in order to
find zero modes of the operators $(\Gamma_X \pm \hat\Gamma_X)$.
Our formulation has $SO(3)$ Poincare invariance
on each $S^2$, whose generators are written as 
\beq
(M_i)_X = (L_i - L_i^R + \frac{\sigma_i}{2})_X \ .
\label{eqn:defM}
\eeq
We then consider the eigenstates of 
the Casimir operator $\sum _i(M_i)_X^2$ as
\beq
\sum_i (M_i)^2_X | J_X \rangle = J_X(J_X+1)| J_X \rangle \ .
\label{eigenstate_of_M2}
\eeq
One can show from the $SU(2)$ algebra of (\ref{eqn:defM})
that the spin $J_X$ takes values 
 $J_X = \frac{1}{2}, \frac{3}{2}, \cdots, n-\frac{1}{2}$.
There are some degeneracies in the states $| J_X \rangle$.
In addition to the $(2J_X+1)$-folded degeneracy 
associated with $(M_3)_X$,
the state $| J_X \rangle$ has a two-folded degeneracy
for $J_X = \frac{1}{2}, \frac{3}{2}, \cdots, n-\frac{3}{2}$.
The highest spin state with $J_X = n-\frac{1}{2}$, however,
does not have this two-folded degeneracy.
As we show in detail in the Appendix \ref{sec:Spectrum_Free}, 
we can see that the Dirac
operator $(\Gamma-\hat\Gamma)_X$ on each $S^2$
does not have a zero mode at all in the free case.
On the other hand, the operator
 $(\Gamma+\hat\Gamma)_X$  does have zero 
 modes in  the highest 
 spin states with $J_X = n-\frac{1}{2}$.
[See a comment below (\ref{CJsqrt}).]
One can also show that
$\Gamma_X |J_X = n-\frac{1}{2}\rangle
=-\hat\Gamma_X |J_X = n-\frac{1}{2}\rangle
=-|J_X = n-\frac{1}{2}\rangle$.

Therefore, coming back to the fuzzy $S^2\times S^2$,
the chiral zero modes of the Dirac operator
$D_{\rm GW}$ are given by the highest spin states
with $J_1=J_2=n-\frac{1}{2}$.
The chirality defined by an eigenvalue of
(\ref{eqn:def_G}) and (\ref{eqn:def_HatG})
is $1$ for all of these states.
The degeneracy of these states is 
$(2J_1+1)(2J_2+1)=4n^2$,
which indeed gives the  first term of 
(\ref{ComLimTopCharResult}).

In the commutative limit, the operator 
$(\Gamma+\hat\Gamma)_X$ 
becomes proportional to
the chirality operator on each $S^2$ and does not have
zero modes. 
In the case of the fuzzy $S^2$, the highest spin states 
have nonzero eigenvalues of the GW Dirac operator (\ref{defDGWX})
and
do not contribute to the index.\footnote{
The highest spin states
have zero eigenvalues of 
the Dirac operator with exact chirality 
 \cite{Carow-Watamura:1996wg},
but have nonzero eigenvalues of the Dirac operator
introduced in \cite{Grosse:1994ed}
and the GW Dirac operator (\ref{defDGWX}).
}
In fuzzy $S^2\times S^2$, however, as we have shown above,
these states become  zero modes of 
 the Dirac operator (\ref{eqn:DGW}) since it 
contains the operator $(\Gamma + \hat\Gamma)_X$.
This is the reason why, even for the free case, 
there is a nonvanishing
term in the topological charge defined 
by the Dirac operator.

\subsection{Monopole configurations and chiral zero modes}
\label{sec:monopole}

In this section, we consider a monopole configuration
as topologically nontrivial gauge field configurations.
We also introduce a modified index theorem and a
topological charge that gives nonvanishing values
for such configurations.
We then investigate the chiral zero modes of 
the GW Dirac operator in these backgrounds.

In the case of the fuzzy $S^2$,
we constructed a 't Hooft-Polyakov monopole configuration 
where the gauge symmetry  group $U(2)$ is 
spontaneously-broken down to $U(1) \times U(1)$
\cite{Balachandran:2003ay, AIN3, AIM, Aoki:2008qta}.
Since the diagonal $U(1)$ is decoupled in the commutative limit,
we discuss only the $SU(2)$ part of the gauge group in the following.
With the $SU(2)$ gauge group broken down to $U(1)$,
this configuration is interpreted as
the 't Hooft-Polyakov type monopole containing both of the 
scalar field with a nonvanishing vev and the monopole gauge field 
configuration on $S^2$.

Analogously,
we now consider $U(2) \times U(2)$ gauge theory on fuzzy $S^2 \times S^2$.
In the presence of the monopole configuration,
the gauge symmetry is spontaneously broken from 
$SU(2) \times SU(2)$ to $U(1) \times U(1)$.
The monopole configuration we will investigate is the following:
\beqa
 (A_a)_1 &=& L_a \otimes \mbox{\boldmath  $1$}_2  
\otimes \mbox{\boldmath  $1$}_2
+ \mbox{\boldmath  $1$}_n \otimes \frac{\tau_a}{2}
\otimes \mbox{\boldmath  $1$}_2 
\doteq  \left(
\begin{array}{cc}
L_a^{(n+1)}  &    \\
 & L_a^{(n-1)}        
\end{array}
\right) \otimes \mbox{\boldmath  $1$}_2, \nonumber \\
\label{eqn:defMonoAa} \\ 
(A_i)_2   &=& L_i \otimes  \mbox{\boldmath  $1$}_2
  \otimes \mbox{\boldmath  $1$}_2
+  \mbox{\boldmath  $1$}_n 
\otimes  \mbox{\boldmath  $1$}_2 \otimes \frac{\tau_i}{2}
\label{eqn:defMonoAi}
\eeqa
where $(A_a)_1$ and $(A_i)_2$ are
covariant coordinates
of the first and the second sphere.
The second and the third factors in the tensor product  
refer to spin $1/2$ representation 
of each $SU(2)$ in the $SU(2)\times SU(2)$ gauge group
respectively.
The equality $\doteq$ means 
a unitary equivalence, and we have combined the first
two spaces, {\it i.e.}, matrix space representing the coordinates
and the first $SU(2)$ space,
into a single matrix representation.
$(A_i)_2$ can be similarly written.
Each of the configurations 
describes the 't Hooft-Polyakov type monopole on each
$S^2$,  and wraps  around the $S^2$.
The normal components of the gauge fields, which are
interpreted as two scalar fields on $S^2 \times S^2$,
have nonvanishing vev's and break the gauge symmetry.

More generally, we can consider the following 
type of configurations:
\beq
 (A_a)_1 = \left(
\begin{array}{cc}
L_a^{(n+m_1)}  &    \\
 & L_a^{(n-m_1)}        
\end{array}
\right) \otimes  \mbox{\boldmath  $1$}_2.
\label{eqn:defMonoA}
\eeq
A generalized $(A_i)_2$ can be written similarly.
For such configurations, the relation
$[(A_a)_1, (A_i)_2]=0$ is satisfied.

Although they are the noncommutative analogs of 
topologically nontrivial configurations,
the topological charge defined in (\ref{eqn:indextheogen})
vanishes for these configurations.
This can be understood as follows:
In the presence of the monopole configurations,
the gauge group is spontaneously broken from $SU(2) \times SU(2)$
to $U(1) \times U(1)$.
A fermionic field in the fundamental representation of each
$SU(2)$ is decomposed into two fermions with 
opposite electric charges $\pm1/2$ of
each of the unbroken $U(1)$'s, and they cancel the topological
charge, or the index of the Dirac operator.

We thus have to modify the index theorem (\ref{eqn:indextheogen})
to pick up one of the fermions with $\pm1/2$ electric charges.
As is shown in section II D of ref.\cite{AIM},
we can prove the following index theorem 
in the projected space:
\beq
\mbox{index}
\left(P_1^{(n\pm m_1)}P_2^{(n \pm m_2)} D_{\rm GW} \right)
=
\frac{1}{2} \CTr \left[
P_1^{(n \pm m_1)}P_2^{(n \pm m_2)} 
(\Gamma + \hat\Gamma) 
\right]  \ ,
\label{indextheorem_U2U2monopole}
\eeq
where $P_X^{(n \pm m_X)}$ is the projection operator 
on the Hilbert space with $n\pm m_X$ dimensions 
in (\ref{eqn:defMonoA}).
The projection operator is written as  
\beq
P_X^{(n \pm m_X)} = \frac{1}{2} (1\pm T_X) \ , 
\eeq
with
\beq
T_X = \frac{2}{n m_X} \left(
(A_X)^2 -\frac{n^2 + m_X^2 -1}{4}
\right) =
\left(
\begin{array}{cc}
\mbox{\boldmath  $1$}_{n+m_X}  &    \\
 & -\mbox{\boldmath  $1$}_{n-m_X}        
\end{array}
\right) \ .
\label{defTX}
\eeq
Here we have left out the extra $\mbox{\boldmath  $1$}_2$.
The operator $T_X$ is interpreted as an electric charge operator
of the unbroken $U(1)$ gauge group.
Its commutative limit becomes the normalized scalar field 
as
\beq
T_X \to 2\phi'_X \ ,
\eeq
where $\phi'_X = \phi^{\prime a}_X \frac{\tau^a}{2}$
with $\sum_a (\phi^{\prime a}_X)^2 = 1$.
Without loss of generality,
we hereafter consider only the following projection:
\beq
P_X^{(n+m_X)} \equiv P_X
\eeq
with $m_X >0$.

Following the same calculation that led us to 
(\ref{ComLimTopCharResult}) in section 
\ref{sec:com_lim_top_char},
the commutative limit of the rhs of 
(\ref{indextheorem_U2U2monopole}) becomes
\beqa
&& \frac{1}{2}\CTr\bigl[P_1 P_2 (\Gamma + \hat\Gamma)\bigr] 
\ \to \ 4(n+m_1)(n+m_2) \n
&&+2(n+m_1)\rho^2 
\int \frac{d\Omega_2}{4\pi}
\epsilon_{ijk} n_{k} \, \tr_2(\phi'_2F_{ij}) 
+2(n+m_2)\rho^2 
\int \frac{d\Omega_1}{4\pi} 
\epsilon_{abc} n_{c} \, \tr_1(\phi'_1F_{ab}) \n
&&+2\rho^4 
\int \frac{d\Omega_1}{4\pi} \frac{d\Omega_2}{4\pi}
\epsilon_{abc} n_{c}\epsilon_{ijk} n_{k} \,
\tr_1(\phi'_1F_{ab}) \tr_2(\phi'_2F_{ij}) \ , 
\label{ComLimTopCharU2U2monopole}
\eeqa
where the $\tr_X$ stands for the trace over the
$SU(2)_X$ gauge group.
The monopole configuration (\ref{eqn:defMonoA})
has the monopole number $(-m_X)$, and  
its 1st Chern character on each $S^2$ becomes $(-m_X)$, 
as is shown below in (\ref{topcharmonoS2-mX}).
Then, (\ref{ComLimTopCharU2U2monopole}) becomes
\beqa
&&4(n+m_1)(n+m_2)+2(n+m_1)(-m_2)+2(n+m_2)(-m_1)+2(-m_1)(-m_2) \n
&=&4n^2+2n(m_1+m_2)+2m_1m_2 \ .
\label{ComLimTopCharU2U2monopole2}
\eeqa

In the following, we will calculate the both-hand sides of
(\ref{indextheorem_U2U2monopole}) at the matrix level,
{\it i.e.}, before taking the commutative limit, 
and check that the result agrees with 
(\ref{ComLimTopCharU2U2monopole2}).
We then investigate what kind of chiral zero modes contribute to 
each term in (\ref{ComLimTopCharU2U2monopole}).

We first calculate the rhs of 
(\ref{indextheorem_U2U2monopole}).
Because of the relation
$[(A_a)_1, (A_i)_2]=0$,
it can be written as
\beq
\frac{1}{2}\Bigl(\CTr_1[P_1\Gamma_1] \CTr_2[P_2\Gamma_2]
+\CTr_1[P_1\hat\Gamma_1] \CTr_2[P_2\hat\Gamma_2] \Bigr) \ .
\label{t1p1g1_t2p2g2}
\eeq
Each factor can be  evaluated as
\beq
\CTr_X[P_X\Gamma_X]=-2(n+m_X) \ , \ \
\CTr_X[P_X\hat\Gamma_X]=2n \ .
\label{resulttpg}
\eeq 
The operator $\Gamma_X$ takes its eigenvalue $\pm 1$
in $n \mp 1$ dimensional representation space 
of the operator $-L^R_i+\sigma_i/2$.
By counting the total dimensions of the space,
including the space on which $PA_i$ acts,
one obtains the first result.
The second result is similarly obtained. 
[See eqs.(3.34) and (3.36) in \cite{AIN3}.]
Then one can obtain the monopole charge on each $S^2$ as
\beq
\frac{1}{2} \CTr \left[
P_X
(\Gamma_X + \hat\Gamma_X) 
\right]  
=-m_X \ .
\label{topcharmonoS2-mX}
\eeq
Substituting (\ref{resulttpg})
into (\ref{t1p1g1_t2p2g2}),
we obtain
\beq
\frac{1}{2}
\Bigl(\left(-2(n+m_1)\right) \left(-2(n+m_2)\right)
+(2n)^2 \Bigr) \ ,
\label{eqn:resIndexAlge}
\eeq
which indeed agrees with the above calculation 
in the commutative limit 
(\ref{ComLimTopCharU2U2monopole2}).

We next calculate the left-hand side (lhs) of 
(\ref{indextheorem_U2U2monopole})
by counting the chiral zero modes 
of the GW Dirac operator 
in the monopole backgrounds.
The commutativity $[\hat\Gamma_1, \hat\Gamma_2]=0$
holds because of the 
relation $[(A_a)_1, (A_i)_2]=0$.
Then, the chirality operator (\ref{eqn:def_HatG}) reduces to  
$\hat\Gamma = \hat\Gamma_1 \hat\Gamma_2$,
and the GW Dirac operator in the projected space becomes
\beq
P_1 P_2 D_{\rm GW} = P_1 P_2 D_1+P_1 P_2 D_2
\label{GWD_monopolebackground_D1D2}
\eeq
with $D_1$ and $D_2$ given in (\ref{def_D1D2_free}).
The arguments we have given  in the free case 
can be applied to the present case, and 
it is sufficient to investigate the  zero modes 
of the operators $P_X(\Gamma_X - \hat\Gamma_X)$
and $P_X(\Gamma_X + \hat\Gamma_X)$
on each fuzzy $S^2$.

We then classify the states in terms of 
the Casimir operator of the $SO(3)$ Poincare symmetry on each $S^2$.
Generators of the $SO(3)$ symmetry
are given by
\beq
(M_i)_X = (P A_i -L_i^R + \frac{\sigma_i}{2})_X \ ,
\eeq
where $A_i$'s are generalized monopole 
configurations (\ref{eqn:defMonoA}).
We consider eigenstates of the Casimir operator $\sum_i ( M_i)^2$
as in (\ref{eigenstate_of_M2}).
As is shown in detail in section III of ref.\cite{AIM},
in addition to the $(2J_X+1)$-folded degeneracy,
the state $| J_X \rangle$ has an extra two-folded degeneracy
for $J_X = \frac{m+1}{2}, \frac{m+3}{2}, \ldots , n+\frac{m-3}{2}$,
while the lowest spin sate with $J_X = \frac{m-1}{2}$
and the highest spin state with $J_X = n+\frac{m-1}{2}$
do not have such two-folded degeneracy.
The lowest spin states are shown to be  zero modes of
the operator $P_X(\Gamma-\hat\Gamma)_X$,
while the highest spin states are zero modes of 
the operator 
$P_X(\Gamma+\hat\Gamma)_X$.
The other states have nonzero eigenvalues for both of these
operators. 
One can also show that
$\Gamma_X |J_X = \frac{m-1}{2}\rangle
=\hat\Gamma_X |J_X = \frac{m-1}{2}\rangle
=-|J_X = \frac{m-1}{2}\rangle$
and that
$\Gamma_X |J_X = n+\frac{m-1}{2}\rangle
=-\hat\Gamma_X |J_X = n+\frac{m-1}{2}\rangle
=-|J_X = n+\frac{m-1}{2}\rangle$.

Consequently, coming back to the fuzzy $S^2\times S^2$,
the chiral zero modes of the
GW Dirac operator $P_1P_2D_{\rm GW}$
in the monopole background (\ref{eqn:defMonoA})
are given by
the lowest spin states with
$J_1=J_2=\frac{m-1}{2}$
and the highest spin states with
$J_1=J_2=n+\frac{m-1}{2}$.
The chirality defined by an eigenvalue of
(\ref{eqn:def_G}) and (\ref{eqn:def_HatG})
is $1$ for all of these states.
The index of the Dirac operator $P_1P_2D_{\rm GW}$
is, therefore, given by counting the degeneracy of these states as
\beq
m_1m_2 +(2n+m_1)(2n+m_2) \ .
\label{deg_chiralzero_monopole_S2S2}
\eeq
This again agrees with the topological charge in the commutative
limit (\ref{ComLimTopCharU2U2monopole2}).
Incidentally, the states with $J_1=\frac{m-1}{2}$,
$J_2=n+\frac{m-1}{2}$ have nonzero eigenvalues
of the operator $P_1 P_2 D_2$, 
and hence do not give chiral zero modes of 
the Dirac operator $P_1 P_2 D_{\rm GW}$.
Neither do the states with $J_1=n+\frac{m-1}{2}$,
$J_2=\frac{m-1}{2}$ contribute to 
chiral zero modes of the Dirac operator $P_1 P_2 D_{\rm GW}$.

Note that the lowest spin states are responsible for
 the first term of 
(\ref{deg_chiralzero_monopole_S2S2}).
This is half of the last term in the rhs of
(\ref{ComLimTopCharU2U2monopole2}), and exactly
matches with an integral of the 2nd Chern character 
in the monopole background we are considering.
This is  reasonable 
since the lowest spin states 
correspond to the chiral zero modes of the Dirac operator 
in the commutative theory. 
All the other contributions to the zero modes in
(\ref{deg_chiralzero_monopole_S2S2}),
and hence in (\ref{ComLimTopCharU2U2monopole2})
and (\ref{ComLimTopCharU2U2monopole}),
come from the highest spin states, which do not have 
corresponding  chiral zero modes in the commutative theory.
Going back to the formula (\ref{ComLimTopCharResult}),
we can similarly infer the origins of various terms.
%

\section{Generalization to fuzzy $(S^2)^k$}
\label{sec:higherdimension}
\setcounter{equation}{0}

In this section, we generalize our formulation to 
fuzzy $(S^2)^k$.
As in the fuzzy $S^2 \times S^2$,
we first define two chirality operators as
\beqa
\Gamma &=& \Gamma_1\cdots\Gamma_k  \ , \\
\hat\Gamma &=&
\frac{\hat\Gamma_1 \cdots \hat\Gamma_k
+\hat\Gamma_k \cdots \hat\Gamma_1}
{\sqrt{(\hat\Gamma_1 \cdots \hat\Gamma_k
+\hat\Gamma_k \cdots \hat\Gamma_1)^2}} \ ,
\eeqa
which satisfy (\ref{eqn:Gh_G^2_relation}).
As in (\ref{relanticomcom}),
the denominator is written as
\beq
(\hat\Gamma_1 \cdots \hat\Gamma_k
+\hat\Gamma_k \cdots \hat\Gamma_1)^2
= 4+(\hat\Gamma_1 \cdots \hat\Gamma_k
-\hat\Gamma_k \cdots \hat\Gamma_1)^2 \ .
\eeq
The second term is of order ${\cal O}(n^{-4})$,
since $[\hat\Gamma_X, \hat\Gamma_Y]$ 
is of order ${\cal O}(n^{-2})$
as is shown below (\ref{comhatgam12}). 
We then define a GW Dirac operator as in (\ref{defDGW}).
It satisfies the GW relation (\ref{eqn:GWrelation}) 
and the index theorem (\ref{eqn:indextheogen}).

Analogously to (\ref{eqn:G-Grel}), 
the following relation is satisfied:
\beqa
&&\Gamma_1 \cdots \Gamma_k 
- \hat\Gamma_1 \cdots \hat\Gamma_k \n
&=&\frac{1}{2^{k-1}}
\sum_{n_1, \cdots , n_k=0,1} 
\frac{1}{2}\left(1-(-1)^{\sum_{X=1}^k n_X}\right) 
\prod_{X=1}^k  \left(\Gamma_X + (-1)^{n_X} \hat\Gamma_X\right) \ , 
\label{identity_in_S2k} 
\eeqa
where the product respects the ordering of operators
from $X=1$ to $X=k$.
The coefficient $(1-(-1)^{\sum_{X=1}^k n_X})$ ensures 
the number of operators $(\Gamma-\hat\Gamma)_{X}$ 
in the product to be odd.
Since $\Gamma_X$ and $\hat\Gamma_X$ become the same
chirality operator in the commutative limit,
those terms with smaller number of $(\Gamma-\hat\Gamma)_X$
in (\ref{identity_in_S2k})
are more dominant in the commutative limit.
%

Then, as in (\ref{eqn:DGWcom}), the commutative limit
of the GW Dirac operator (\ref{defDGW}) becomes
\beq
D_{\rm GW} \to
D'_{1} \gamma_2 \cdots \gamma_k 
+ \gamma_1 D'_{2} \gamma_3\cdots\gamma_k 
+\cdots
+ \gamma_1 \cdots\gamma_{k-1} D'_{k} \ ,
\eeq
where only the terms with one of the $n_X$'s being 1
in (\ref{identity_in_S2k}) contribute.
This is a generalized Dirac operator
on a commutative $(S^2)^k$.
[See the discussion after eq. (\ref{eqn:DGWcom}).]

The commutative limit of the topological charge,
the rhs of (\ref{eqn:indextheogen}),
gives a generalization of (\ref{ComLimTopCharResult})
and (\ref{ComLimTopCharResult_form}). 
We now conjecture the result as follows:
\beq
\frac{1}{2} \CTr [\Gamma + \hat\Gamma] \to
(1+(-1)^k)2^{k-1}n^k \tr({\bf 1})
+2^{k-1} \sum_{i=1}^k n^{k-i} C_i \ .
\label{ComLimTopCharS2k}
\eeq
The coefficient $(1+(-1)^k)$ in the first term 
represents that this term vanishes when $k$ is odd.
This is because the contributions of the two chirality 
operators cancel for odd $k$. 
The integral  of the
$i$-th Chern character $C_i$ is defined as
\beq
C_i = \frac{1}{(2\pi)^k i!} \int \tr \left[
\sum_{1\le X_1 < \cdots < X_i \le k}\left(
\prod_{X\notin (X_1\cdots X_i)}\frac{d\Omega_X}{2} \
(F^i)_{X_1\cdots X_i} \right)\right] \ .
\eeq
For instance, $(F)_X$ and $(F^2)_{XY}$ 
are given in (\ref{defFX}) and (\ref{defF2XY}), 
and $(F^3)_{XYZ}$ is written as
\beqa
&&\frac{3!}{2^3}\rho^6 
d\Omega_X d\Omega_Y d\Omega_Z
\Bigl(\epsilon_{abc} n_{c}\epsilon_{ijk} n_{k}\epsilon_{xyz} n_{z}
\bigl(F_{ab}F_{ij}F_{xy} 
-F_{ai}F_{bj}F_{xy} \n
&&+F_{aj}F_{bi}F_{xy}
-F_{ab}F_{ix}F_{jy}+F_{ab}F_{iy}F_{jx}
-F_{ax}F_{by}F_{ij}+F_{ay}F_{bx}F_{ij} \n
&&+F_{ai}F_{bx}F_{jy}-F_{ai}F_{by}F_{jx}
-F_{aj}F_{bx}F_{iy}+F_{aj}F_{by}F_{ix} \n
&&-F_{ax}F_{bi}F_{jy}+F_{ay}F_{bi}F_{jx}
+F_{ax}F_{bj}F_{iy}-F_{ay}F_{bj}F_{ix}\bigr)\Bigr)_{XYZ} \ ,
\label{defF3XYZ}
\eeqa
where the indices $a$, $b$, and $c$ refer to the sphere $X$,
the indices $i$, $j$, and $k$ to the sphere $Y$,
and the indices $x$, $y$, and $z$ to the sphere $Z$.
Note, however, that the field strength depends 
on all of the coordinates,
such as $F_{ab}(\Omega_1, \cdots, \Omega_k)$.
Only the highest Chern character term in 
(\ref{ComLimTopCharS2k}) is independent 
of the size $n$ of the matrix.
It is  important 
to show the conjecture (\ref{ComLimTopCharS2k}) explicitly
by taking the commutative limit as 
we did for the fuzzy $S^2\times S^2$ in
section \ref{sec:com_lim_top_char}.
It needs  involved calculations 
and we will report it in a future publication.

We here demonstrate the justification of (\ref{ComLimTopCharS2k})
by considering a topologically nontrivial configuration, {\it i.e.},
a monopole configuration
in $(SU(2))^k$ gauge theory on fuzzy  $(S^2)^k$.
It is a generalization of (\ref{eqn:defMonoA}).
As in (\ref{indextheorem_U2U2monopole}),
we consider 
the index theorem in the projected space
\beq
\mbox{index}
\left(P_1 \cdots P_k D_{\rm GW} \right)
=
\frac{1}{2} \CTr \left[
P_1 \cdots P_k (\Gamma + \hat\Gamma) 
\right]  \ .
\label{indextheorem_U2kmonopole}
\eeq
If the conjecture (\ref{ComLimTopCharS2k}) holds,
then as in
(\ref{ComLimTopCharU2U2monopole}),
the commutative limit of the rhs of 
(\ref{indextheorem_U2kmonopole}) becomes
\beqa
&&\frac{1}{2} \CTr \left[
P_1 \cdots P_k (\Gamma + \hat\Gamma) 
\right]  
\ \to \ (1+(-1)^k)2^{k-1} \prod_{X=1}^k (n+m_X) \n
&& 
+2^{k-1} \sum_{i=1}^k 
\left[\sum_{1 \le X_1 < \cdots < X_i \le k} \right. \n
&&
\left. \left(
\prod_{X \notin (X_1 \cdots X_i)}(n+m_X)
\prod_{X \in (X_1 \cdots X_i)}
\rho^2 
\left(\int \frac{d\Omega}{4\pi}
\epsilon_{ijk} n_{k} \, \tr(\phi'F_{ij}) \right)_X
\right) \right] \ . \n
\label{comlim_topchar_U2kmon}
\eeqa
The monopole on each $S^2$ gives 
the 1st Chern character $(-m_X)$.
Following the same calculation as in
(\ref{ComLimTopCharU2U2monopole2}),
(\ref{comlim_topchar_U2kmon}) becomes
\beq
(1+(-1)^k)2^{k-1} n^k 
+ (-1)^k2^{k-1}\sum_{i=1}^kn^{k-i}
\sum_{1\le X_1 < \cdots < X_i\le k}m_{X_1} \cdots m_{X_i} \ .
\label{topchar_U2kmonopole_S2k}
\eeq

In the following, we will evaluate the both-hand sides
of (\ref{indextheorem_U2kmonopole}) at the matrix level,
{\it i.e.}, before taking the commutative limit,
and show that the results agree with 
the conjectured  topological charge in the commutative
limit (\ref{topchar_U2kmonopole_S2k}).

Following the same calculations in (\ref{t1p1g1_t2p2g2})
and (\ref{eqn:resIndexAlge}),
the rhs of (\ref{indextheorem_U2kmonopole}) 
for the monopole background becomes
\beqa
&&\frac{1}{2} \left(
\prod_{X=1}^k \CTr_X[P_X\Gamma_X] 
+ \prod_{X=1}^k \CTr_X[P_X\hat\Gamma_X] 
 \right)  \n
&=&\frac{1}{2} \left(
\prod_{X=1}^k \bigl( -2(n+m_X) \bigr) + (2n)^k \right) \ ,
\label{topchar_monoS2k}
\eeqa
which indeed gives (\ref{topchar_U2kmonopole_S2k}).

We can also evaluate the lhs of (\ref{indextheorem_U2kmonopole})
by counting the chiral zero modes of the Dirac operator.
Denoting each term in (\ref{identity_in_S2k})
as $D_a$ with $a=1, \ldots, 2^{k-1}$,
we obtain a generalization of 
eq. (\ref{GWD_monopolebackground_D1D2}).
The same arguments we have given in the $S^2\times S^2$ case 
hold in the present case:
A chiral zero mode of the Dirac operator 
$P_1 \cdots P_k D_{\rm GW}$ must be
a simultaneous zero mode of all the operators 
$P_1 \cdots P_k D_a$
with $a=1, \ldots, 2^{k-1}$.
A zero mode of $P_1 \cdots P_kD_a$ is given by  
a zero mode of any of the operators $P_X(\Gamma +\hat\Gamma)_X$
and $P_X(\Gamma -\hat\Gamma)_X$
constituting $P_1 \cdots P_kD_a$.
The lowest spin states with $J_X=\frac{m-1}{2}$
are zero modes of the operator $P_X(\Gamma -\hat\Gamma)_X$,
and the highest spin states with $J_X=n+\frac{m-1}{2}$
are zero modes of the operator $P_X(\Gamma +\hat\Gamma)_X$.
Eventually, we find that the chiral zero modes of 
the Dirac operator
$P_1 \cdots P_k D_{\rm GW}$ are given by the states
where an even number of $J_X$'s are the highest spin 
and the remaining $J_X$'s are the lowest spin. 
The chirality defined by an eigenvalue of (\ref{eqn:def_G}) 
and (\ref{eqn:def_HatG})
is $1$ for all of these states when $k$ is even,
and $-1$ when $k$ is odd.
By counting the number of these states 
as in (\ref{deg_chiralzero_monopole_S2S2}),
the index of the Dirac operator 
$P_1 \cdots P_k D_{\rm GW}$ is evaluated as
\beq
(-1)^k \sum_{i=0,2,\cdots}  \left[
\sum_{1 \le X_1 < \cdots < X_{i} \le k} 
\left( \prod_{X \in (X_1,\cdots,X_i)}(2n+m_X) 
\prod_{X \notin (X_1,\cdots,X_i)} m_X \right)\right] \ .
\label{index_monoS2k}
\eeq
This again reproduces the result  
(\ref{topchar_U2kmonopole_S2k}).
Incidentally, the states with an odd number $i$ of 
$J_X$ being the highest spin,
which we call $J_{X_1}, \ldots, J_{X_i}$,
have nonzero eigenvalues of the operator $P_1 \cdots P_kD_a$
that is composed of $(\Gamma-\hat\Gamma)_X$
with $X \in (X_1, \ldots, X_i)$
and $(\Gamma+\hat\Gamma)_X$
with $X \notin (X_1, \ldots, X_i)$.
Those states thus do not contribute to the chiral zero modes
of the Dirac operator $P_1 \cdots P_k D_{\rm GW}$.
We also note that the states with all $J_X$ being the lowest spin
are responsible for the term with $i=0$ in (\ref{index_monoS2k}), 
giving $\prod_{X=1}^k (-m_X)$,
which agrees precisely with the $k$-th Chern character
of the background gauge fields we are considering.
This is reasonable since these states correspond
to the chiral zero modes in the commutative theory.

The agreement of (\ref{topchar_monoS2k}) 
and (\ref{index_monoS2k})
to (\ref{topchar_U2kmonopole_S2k})
supports the conjecture
(\ref{comlim_topchar_U2kmon}),
and hence 
(\ref{ComLimTopCharS2k}).

\section{Conclusions and Discussions}
\label{sec:conclusion}
\setcounter{equation}{0}

In this paper, we have constructed a topological 
charge on the fuzzy $(S^2)^k$
based on a Dirac operator 
satisfying the GW relation. 
Our formulation has the manifest gauge invariance
and the $SO(3)$ Poincare invariance on each $S^2$.
Owing to the GW relation, the index theorem is satisfied
and accordingly we can construct the topological charge.
The commutative limit
of the topological charge 
was evaluated directly for the fuzzy $S^2\times S^2$,
and it becomes a sum of the 1st and the 2nd Chern characters.
We then have shown that by combining with other topological
invariants we can define a noncommutative
generalization of the 2nd Chern character.
We also conjectured a form of 
the commutative limit of the topological charge
on  fuzzy $(S^2)^k$ for $k>2$.

We further calculated the chiral zero modes of
the Dirac operator for the free case and
for the monopole backgrounds,
and checked the consistency of our results.
The zero modes of the noncommutative GW Dirac operator
on fuzzy $(S^2)^k$
consist of the highest spin states and the lowest spin states.
The lowest spin states correspond to the zero modes of the
commutative Dirac operator. On the other hand,
the highest spin states are zero modes of the  operator
$(\Gamma + \hat\Gamma)_X$ and 
do not have the correspondents in the commutative limit. 
We have indeed found that
the chiral zero modes
composed of only the lowest spin states
give precisely the $k$-th Chern character on $(S^2)^k$.


Some comments are in order.
 In the definition of $\hat\Gamma$ in (\ref{eqn:def_HatG}),
 we first normalized both of $\hat\Gamma_X$ in (\ref{def_hatGforS2}),
 and then 
 constructed the normalized chirality operator $\hat\Gamma$
 on $S^2 \times S^2$ in (\ref{eqn:def_HatG}).
 Instead, we can directly construct a normalized  operator
 on $S^2 \times S^2$ as
\beq
\hat\Gamma' = \frac{\{ H_1, \ H_2 \}}{\sqrt{\{ H_1, \ H_2 \}^2}} \ ,
\label{defhatgamma_onetime}
\eeq
with $H_X$ defined in (\ref{def_hatGforS2}). 
Defining a Dirac operator as in (\ref{defDGW}),
with $\hat\Gamma$ replaced by $\hat\Gamma'$,
the GW relation 
(\ref{eqn:GWrelation}) 
and the index theorem (\ref{eqn:indextheogen})
are satisfied as well.
Moreover, as we show in 
Appendix \ref{sec:one_time_normalization},
the commutative limit of the Dirac operator
and the topological charge give exactly the same result as 
(\ref{eqn:DGWcom}) and (\ref{ComLimTopCharResult}).
This agreement indicates
that the topological quantities are rigid against 
slight modifications of the theories.

In this paper, we considered the monopole configurations
wrapping around each $S^2$, but
it is more interesting if we can construct  configurations 
wrapping around higher dimensional space.
Then the field strengths whose indices mix the different spheres
play an important role. 
It is also interesting, as we have studied
for the case of fuzzy $S^2$ in ref.\cite{Aoki:2008qta},
to further extend our formulation
of the projected index theorem to include more
general configurations in the Higgs phase,
{\it i.e.}, when the scalar field takes a nonzero vev.

As we mentioned at the beginning of the Introduction,
topological aspects of gauge theory on
noncommutative geometry may play an important role
in compactified extra dimensional space
in string theory.
We can pursue these studies further
by studying the relation of noncommutative geometry to our world 
and by investigating dynamics of noncommutative gauge theory.
(See also related works 
\cite{AIMN,Steinacker:2007ay,Aoki:2006sb}.)
Our formulation given in the present paper 
to define the topological charge
and to classify the gauge field configuration space
on noncommutative geometry 
will become useful for these studies.


\appendix

\section{Expansion of $\hat\Gamma$ in the gauge fields}
\label{sec:calOfTrG}
\setcounter{equation}{0}

In this appendix, we 
expand the chirality operator $\hat\Gamma$ in terms of 
the gauge fields, 
and provide (\ref{hatG_exp_n-1_-4}).

We first expand the chirality operator $\hat\Gamma_X$ 
on each $S^2$, defined by (\ref{def_hatGforS2}).
We decompose $H_X$ into the 0-th  
and the 1st order in the gauge fields as 
\beq
H_X = \al_X +\be_X \ ,
\label{defH_albe}
\eeq 
with 
\beq
\al_X = a\left( \sigma_i L_i + \frac{1}{2} \right)_X \ , \ \ \
\be_X = a\rho (\sigma_i a_i)_X \ .
\label{eqn:def_al_be}
\eeq
The operators $\al_X$ and $\be_X$ are
of order ${\cal O}(n^0)$ and ${\cal O}(n^{-1})$,
respectively, 
since $a=2/n$ and $L_i$ is  of order $n$.
Since $(\al_X)^2=1$, one has
$(H_X)^2 = 1+ \{ \al_X, \ \be_X \} +\be_X^2$.
We then obtain
\beq
\hat\Gamma_X
=\left(\al+\zeta^{(1)}+\zeta^{(2)}
+\zeta^{(3)}+{\cal O}(\be^4)\right)_X
\label{eqn:hatGxExpand}
\eeq
where $\zeta^{(i)}_X$ is the $i$-th order in $\be_X$
and hence
in the gauge field $(a_i)_X$.
They are written as
\beqa
\zeta^{(1)}_X&=&\frac{1}{2} (\be - \al \be \al)_X \ , 
\label{hatGX_exp_a1}\\
\zeta^{(2)}_X&=&\left(-\frac{1}{8} (\al \be^2 +\be \al \be + \be^2 \al) 
+\frac{3}{8}\al \be \al \be \al \right)_X \ ,
\label{hatGX_exp_a2}\\
\zeta^{(3)}_X&=& \left(\frac{1}{16}
(-\be^3+\be\al\be\al\be +\be\al\be^2\al +\be^2\al\be\al
+\al\be\al\be^2 +\al\be^2\al\be +\al\be^3\al) \right.\n
&&\left. -\frac{5}{16}\al\be\al\be\al\be\al \right)_X \ ,
\label{hatGX_exp_a3}
\eeqa
The operators $\al_X$ and $\zeta^{(i)}_X$ themselves
are $0$-th and $i$-th order
in $1/n$.
However,
taking the trace over the spinor space
with the coordinate matrix space untouched,
the operators
$\tr_{\sigma_X}(\al_X)$ and $\tr_{\sigma_X}(\zeta^{(1)}_X)$
become of order $n^{-1}$ and $n^{-2}$, respectively.

It then follows that 
\beqa
\{ \hat\Gamma_1, \ \hat\Gamma_2 \} 
&=& 2\al_1 \al_2  \n
&&
+\{\al_1, \ \zeta^{(1)}_2+\zeta^{(2)}_2+\zeta^{(3)}_2 \} 
+\{\al_2, \ \zeta^{(1)}_1+\zeta^{(2)}_1+\zeta^{(3)}_1 \} \n
&&
+\{ \zeta^{(1)}_1+\zeta^{(2)}_1, \ \zeta^{(1)}_2+\zeta^{(2)}_2 \} \n
&& +{\cal O}(n^{-5}) \ .
\label{eqn:numHatG_all}  
\eeqa
While the operators $\{\al_1, \ \zeta^{(4)}_2 \}$,
$\{\al_2, \ \zeta^{(4)}_1 \}$,
$\{\zeta^{(1)}_1, \ \zeta^{(3)}_2 \}$
and $\{\zeta^{(1)}_2, \ \zeta^{(3)}_1 \}$
also appear at order $n^{-4}$,
when one considers these terms in $\CTr [\hat\Gamma]$
in (\ref{hatG_exp_n-1_-4}),
one takes a trace like
$\tr_{\sigma_X}(\al_X)$ and $\tr_{\sigma_X}(\zeta^{(1)}_X)$,
and these terms become of order ${\cal O}(n^{-5})$.
One also has
\beq
[ \hat\Gamma_1, \ \hat\Gamma_2 ]
= [\al_1, \ \zeta^{(1)}_2 ]-[\al_2, \ \zeta^{(1)}_1 ]
+[\zeta^{(1)}_1, \ \zeta^{(1)}_2 ]
+{\cal O}(n^{-3}) \ .
\label{comhatgam12}
\eeq
Note that (\ref{comhatgam12})
is of order ${\cal O}(n^{-2})$,
since the leading term $[\al_1,\al_2]$ vanishes,
and the commutators $[\al_1, \ \be_2]$ and
$[\al_2, \ \be_1]$ are of order ${\cal O}(n^{-2})$.
This is why the second term in (\ref{relanticomcom})
is of order ${\cal O}(n^{-4})$.

Using the identity (\ref{relanticomcom}),
the chirality operator (\ref{eqn:def_HatG}) is written as
\beq
\hat\Gamma = \frac{1}{2} \{ \hat\Gamma_1, \ \hat\Gamma_2 \}
-\frac{1}{16} \{ \hat\Gamma_1, \ \hat\Gamma_2 \}
[ \hat\Gamma_1, \ \hat\Gamma_2 ]^2 + \cdots \ .
\label{eqn:TrHatG}
\eeq
Plugging (\ref{eqn:numHatG_all}) and (\ref{comhatgam12})
into (\ref{eqn:TrHatG}),
we obtain (\ref{hatG_exp_n-1_-4}).

\section{Commutative limit of $\CTr [G_5]$}
\label{sec:cal2ndChern}
\setcounter{equation}{0}

In this appendix, we show the equation (\ref{eqn:2ndChern_2})
by taking the commutative limit of 
$\CTr [G_5]$. 
Substituting (\ref{hatGX_exp_a1})
into (\ref{eqn:TrHatG_2_2ndChern2}),
we obtain
\beq
G_5 = \sum_{i=1}^{5} K_i
\eeq
with
\beqa
K_1 &=& -\frac{1}{32} \al_1 \al_2 
\left([\al_1,\ \be_2 ]-[\al_2,\ \be_1 ] \right)^2 \ ,
\label{eqn:TrHatG_2_2ndChern22} \\
K_2 &=& -\frac{1}{32}\al_1 \al_2
  \left( \al_2[\al_1,\ \be_2 ]\al_2 - 
  \al_1[\al_2,\ \be_1 ]\al_1 \right)^2 \ ,
 \label{eqn:TrHatG_2_2ndChern3} \\
K_3 &=& \frac{1}{32}\al_1 \al_2
  \left\{[\al_1,\ \be_2 ]-[\al_2,\ \be_1 ], \
 \al_2[\al_1,\ \be_2 ]\al_2 
 - \al_1[\al_2,\ \be_1 ]\al_1 \right\} \ , 
 \label{eqn:TrHatG_2_2ndChern4} \\
K_4 &=& -\frac{1}{64}\al_1\al_2
\left([\al_1, \ \be_2]-[\al_2, \ \be_1]\right)
[\be_1, \ \be_2] 
+{\rm (15 \ terms)} \ , 
\label{eqn:TrHatG_2_2ndChern_a3} \\
K_5 &=& -\frac{1}{128} \al_1 \al_2
[\be_1, \ \be_2]^2
+{\rm (15 \ terms)} \ ,
\label{eqn:TrHatG_2_2ndChern_a4}
\eeqa
where $K_1$, $K_2$, and $K_3$ are  2nd order,
$K_4$ is 3rd order,
and $K_5$ is 4th order in $\beta$.
In (\ref{eqn:TrHatG_2_2ndChern_a3}) and 
(\ref{eqn:TrHatG_2_2ndChern_a4}),
we wrote only a typical term.
The remaining 15 terms can be similarly written.

We first calculate the commutative limit 
of $\CTr[K_1]$.
Plugging (\ref{eqn:def_al_be}) into (\ref{eqn:TrHatG_2_2ndChern22}),
we obtain
\beqa
\CTr[K_1]
&=& -\frac{1}{32} a^6 \rho^2 \ 
\CTr \Bigr[ (\sigma \cdot L)_1(\sigma \cdot L)_2 
\left[ (\sigma \cdot L)_1, \ (\sigma \cdot a)_2  \right]^2 \n
&&-(\sigma \cdot L)_1(\sigma \cdot L)_2
\left[ (\sigma \cdot L)_1, \ (\sigma \cdot a)_2  \right]
\left[ (\sigma \cdot L)_2, \ (\sigma \cdot a)_1  \right] \n
&&+( 1 \leftrightarrow 2 ) \Bigl] \ ,
\label{eqn:TrHatG_2_2ndChern2_nc1}
\eeqa
where we omitted subleading terms in $1/n$.
Taking trace over the spinor space, 
by using the formula
\beq
\tr_\sigma [ \sigma_i \sigma_j \sigma_k]
= 2 i \epsilon_{ijk} \ , 
\label{eqn:sigma3}
\eeq
(\ref{eqn:TrHatG_2_2ndChern2_nc1}) becomes 
\beqa
&& \frac{1}{8} a^6 \rho^2 \ 
\CTr' \Bigr[ \epsilon_{abc} L_c \epsilon_{ijk} L_k
\left[  L_a, \ a_i  \right] \left[  L_b, \ a_j  \right] \n
&&-\epsilon_{abc} L_c \epsilon_{ijk} L_k
\left[  L_a, \ a_i  \right] \left[  L_j, \ a_b  \right]
+( 1 \leftrightarrow 2 )\Bigl] \ ,
\label{eqn:TrHatG_2_2ndChern2_nc2}
\eeqa
where $\CTr'$ is the trace over the matrix space
and the gauge group space.
The indices $a$, $b$, and $c$ refer to the first $S^2$,
while the indices $i$, $j$, and $k$  refer to the second $S^2$.
Then, the commutative limit 
 of (\ref{eqn:TrHatG_2_2ndChern2_nc2}) becomes 
\beq
-2\rho^4
\int \frac{d\Omega_1}{4\pi} \frac{d\Omega_2}{4\pi}
\tr \left[ \epsilon_{abc} n_{c}\epsilon_{ijk} n_{k}
(\del_a a_i \del_b a_j  + \del_i a_a \del_j a_b )
+2\del_a (Pa)_i  \del_i (Pa)_a     
  \right]  \ ,
  \label{EpsEpsPaPa}
\eeq
where $(Pa)_i = P_{ij} a_j$
with $P_{ij}=\delta_{ij}-n_i n_j$.
(\ref{EpsEpsPaPa})
is rewritten as
\beq
 -2\rho^4 
\int \frac{d\Omega_1}{4\pi} \frac{d\Omega_2}{4\pi}
\tr \left[\epsilon_{abc} n_{c}\epsilon_{ijk} n_{k}
(\del_a a'_i- \del_i a'_a)(\del_b a'_j- \del_j a'_b) \right] \ ,
\label{eeFF}
\eeq
where $a'_i = \epsilon_{ijk} n_j a_k$ is the tangential
component of the gauge field.

By using the identity
\beqa
\al_1[\al_1, \be_2] &=&-[\al_1, \be_2] \al_1 \ , \\
\al_2[\al_2, \be_1] &=&-[\al_2, \be_1] \al_2 \ , 
\eeqa
(\ref{eqn:TrHatG_2_2ndChern3}) is rewritten as
\beq
K_2=-\frac{1}{32}
([\al_1,\ \be_2 ]-[\al_2,\ \be_1 ])^2 \al_1 \al_2 \ ,
\label{eqn:TrHatG_2_2ndChern3rewr}
\eeq
and (\ref{eqn:TrHatG_2_2ndChern4}) is 
\beqa
K_3&=&-\frac{1}{32} \Bigl(
\al_2[\al_1,\ \be_2]\al_1\al_2[\al_1,\ \be_2]\al_2
+[\al_1,\ \be_2]\al_1\al_2[\al_1,\ \be_2] \n
&&
-\al_2[\al_1,\ \be_2][\al_2,\ \be_1]\al_1
-[\al_1,\ \be_2]\al_1\al_2[\al_2,\ \be_1]
+(1 \leftrightarrow 2)\Bigr) \ . 
\eeqa
By the same calculation that was done for
$\CTr[K_1]$,
we can show that the commutative limits of 
$\CTr[K_2]$ and
$\CTr[K_3]$ give
the same result (\ref{eeFF})
and twice of that, respectively.
Therefore, the commutative limit of $\CTr(K_1+K_2+K_3)$
becomes 4 times of (\ref{eeFF}).
This gives the 2nd order terms in the gauge field
in (\ref{eqn:2ndChern_2}).

We next consider $\CTr[K_4]$.
By substituting (\ref{eqn:def_al_be})
and taking the trace over the spinor space,
the first term in $K_4$, which was presented in 
(\ref{eqn:TrHatG_2_2ndChern_a3}),
gives
\beq
\frac{1}{16} a^6 \rho^3 \CTr'
\Bigl[\epsilon_{abc} L_c \epsilon_{ijk} L_k
\left([L_a, \ a_i]-[L_i, \ a_a]\right)[a_b, \ a_j] \Bigr] \ .
\eeq
Its commutative limit becomes
\beq
-i \rho^4 \int \frac{d\Omega_1}{4\pi} \frac{d\Omega_2}{4\pi}
\epsilon_{abc} n_c \epsilon_{ijk} n_k
\tr \bigl(
(\epsilon_{ade}n_d \del_e a_i
-\epsilon_{ilm}n_l \del_m a_a )
[a_b, \ a_j] \bigr) \ .
\eeq
This is rewritten as
\beq
i \rho^4 \int \frac{d\Omega_1}{4\pi} \frac{d\Omega_2}{4\pi}
\epsilon_{abc} n_c \epsilon_{ijk} n_k
\tr\bigl( (\del_a a'_i-\del_i a'_a )
[a'_b, \ a'_j] \bigr) \ .
\label{eeFF_a3}
\eeq
The remaining 15 terms in (\ref{eqn:TrHatG_2_2ndChern_a3})
give the same results.
Thus, the commutative limit of $\CTr[K_4]$ 
becomes 16 times of (\ref{eeFF_a3}).
This gives the 3rd order terms in the gauge field
in (\ref{eqn:2ndChern_2}).

We finally consider $\CTr[K_5]$.
By substituting (\ref{eqn:def_al_be})
and taking the trace over the spinor space,
the first term in $K_5$, which was presented in 
(\ref{eqn:TrHatG_2_2ndChern_a4}),
gives
\beq
\frac{1}{32} a^6 \rho^4 \CTr'
\Bigl[\epsilon_{abc} L_c \epsilon_{ijk} L_k
[a_a, \ a_i] [a_b, \ a_j]\Bigr] \ .
\eeq
Its commutative limit becomes
\beq
\frac{1}{2} \rho^4 \int \frac{d\Omega_1}{4\pi} \frac{d\Omega_2}{4\pi}
\epsilon_{abc} n_c \epsilon_{ijk} n_k
\tr \bigl( [a_a, \ a_i][a_b, \ a_j] \bigr) \ ,
\eeq
which is rewritten as
\beq
\frac{1}{2} \rho^4 \int \frac{d\Omega_1}{4\pi} \frac{d\Omega_2}{4\pi}
\epsilon_{abc} n_c \epsilon_{ijk} n_k
\tr \bigl( [a'_a, \ a'_i][a'_b, \ a'_j] \bigr) \ .
\label{eeFF_a4}
\eeq
The remaining 15 terms in (\ref{eqn:TrHatG_2_2ndChern_a4})
give the same results.
Thus, the commutative limit of $\CTr[K_5]$ 
becomes 16 times of (\ref{eeFF_a4}).
This gives the 4th order terms in the gauge field
in (\ref{eqn:2ndChern_2}).

Hence we have proved (\ref{eqn:2ndChern_2}).


\section{Spectrum of the Dirac operator for the free case}
\label{sec:Spectrum_Free}
\setcounter{equation}{0}


In this appendix, we calculate the whole spectrum of 
the GW Dirac operator for the free case.
We here consider the $U(1)$ gauge group, for simplicity.
For the free case, one has
\beqa
(\Gamma -\hat\Gamma)_X &=& -a( \sigma \cdot \tilde{L} +1)_X \ , \\
(\Gamma +\hat\Gamma)_X &=& a( \sigma \cdot (L+L^R))_X \ ,
\eeqa
where $(\tilde L_i)_X = (L_i - L^R_i)_X$ 
is the adjoint operator.
Then, the free GW Dirac operator (\ref{eqn:DGW}) is written as
\beq
D_{\rm GW} = \frac{a}{2}
\left[( \sigma \cdot \tilde{L} +1)_1( \sigma \cdot (L+L^R))_2
+( \sigma \cdot (L+L^R))_1( \sigma \cdot \tilde{L} +1)_2 
\right] \ .
\label{eqn:DGWfree}
\eeq

Let us begin with an investigation of each fuzzy $S^2$.
Our formulation has $SO(3)$ Poincare symmetry on 
each $S^2$,
whose generator $(M_i)_X$ 
is given in (\ref{eqn:defM}).
We now write its eigenstates as
\beq
(M_i)^2_X | J_X, \pm \rangle = J_X(J_X+1)| J_X, \pm \rangle \ .
\eeq
Each $| J_X, \pm \rangle$ has $(2J_X+1)$-folded degeneracy
associated with $(M_3)_X$.
The sign $\pm$ indicates that this state is obtained from
the spin $l_X$ state of $(\tilde{L}_i)_X$ as
$J_X=l_X \pm \frac{1}{2}$.
For $J_X = \frac{1}{2}, \frac{3}{2}, \ldots, n-\frac{3}{2}$,
there exist both $| J_X, + \rangle$ and $| J_X, - \rangle$,
while for the highest spin $J_X=n-\frac{1}{2}$
there exists only $| J_X, + \rangle$.
The state $| J_X, \pm \rangle$ is shown to be
 an eigenstate of
the operator $(\sigma\cdot \tilde{L} +1)_X$ as 
\beq
(\sigma\cdot \tilde{L} +1)_X | J_X, \pm \rangle 
= \pm \left( J_X+\frac{1}{2} \right) | J_X, \pm \rangle \ . 
\label{eigenstate_of _Dirac_op_free}
\eeq
Since we have the relation
\beq
\{\Gamma -\hat\Gamma , \ \Gamma +\hat\Gamma\}_X=0 \ ,
\eeq
and, in particular for the free case,
\beq
\bigl\{  \sigma \cdot \tilde{L} +1, 
\ \sigma \cdot (L+L^R) \bigr\}_X = 0 \ ,
\label{chiralrelfree}
\eeq
the operator $(\sigma\cdot(L+L^R))_X$
flips the $\pm$ sign as
\beq
(\sigma\cdot(L+L^R))_X| J_X, \pm \rangle 
=C_{J_X} | J_X, \mp \rangle 
 \label{eqn:sLLR_1sphere}
\eeq
with
\beq
C_{J_X}=\sqrt{n^2-\frac{1}{4} -J_X(J_X+1)} \ .
\label{CJsqrt}
\eeq
For the highest spin $J_X = n-\frac{1}{2}$,
a state $| J_X, - \rangle$ does not exist,
and thus $(\sigma\cdot(L+L^R))_X| J_X, + \rangle$ must vanish.
Indeed, $C_{J_X}=0$ in this case,
as one can see from (\ref{CJsqrt}).

We now come back to $S^2 \times S^2$.
We consider states specified by the spin
$J_1$ and $J_2$ of each $S^2$.
We will study the following three cases in turn:
\beq
\begin{array}{l}
(a) \ \displaystyle{ \frac{1}{2} \le J_1 \leq n-\frac{3}{2}, \ 
\frac{1}{2} \leq J_2 \leq n-\frac{3}{2} } \\
(b) \  \displaystyle{ J_1 = n-\frac{1}{2}, \  
\frac{1}{2} \leq J_2 \leq n-\frac{3}{2} } \\
(c) \ \displaystyle{ J_1 =
J_2 = n-\frac{1}{2} }
\end{array}
\eeq

Let us first consider the case (a),
where four types of states $|J_1, \pm ;J_2, \pm \rangle$
exist.
Acting the GW Dirac operator (\ref{eqn:DGWfree})
on these states, we obtain
\beqa
&&D_{\rm GW} \Bigl( c_1 | J_1, +; J_2, + \rangle 
+ c_2 | J_1, +; J_2, - \rangle
+c_3 | J_1, -; J_2, + \rangle  
+c_4 |J_1, -; J_2, - \rangle \Bigr) \n
&=&
 (Ac_2 + Bc_3)| J_1, +; J_2, + \rangle
+(Ac_1 - Bc_4)| J_1, +; J_2, - \rangle \n
&&+(-Ac_4 + Bc_1)| J_1, -; J_2, + \rangle
+(-Ac_3 - Bc_2)|J_1, -; J_2, - \rangle \ ,
\eeqa
with
$A=\frac{a}{2}(J_1+\frac{1}{2})C_{J_2}$ and 
$B=\frac{a}{2}(J_2+\frac{1}{2})C_{J_1}$. 
Diagonalizing $D_{\rm GW}$ in this sector,
we obtain the eigenvalues $\pm |A\pm B|$,
where two $\pm$ signs need not coincide.

In particular, for $J_1 = J_2$, and hence for $A=B$,
there exist two types of zero modes.
Their explicit form is given as
\beqa
&& |1 \rangle 
= \frac{1}{2} \Bigl( | J_1, +; J_2, + \rangle
+| J_1, +; J_2, - \rangle -| J_1, -; J_2, + \rangle
+| J_1, -; J_2, - \rangle \Bigr) \ , \n
&&\\
&&  |2 \rangle 
= \frac{1}{2} \Bigl( | J_1, +; J_2, + \rangle
-| J_1, +; J_2, - \rangle +| J_1, -; J_2, + \rangle
+| J_1, -; J_2, - \rangle \Bigr) \ . \n
&&
\eeqa
We now study their chiralities.
The chirality operator (\ref{eqn:def_G})
is rewritten as
\beqa
\Gamma 
&=& \frac{a^2}{4} \left[ 
(\sigma\cdot (L+L^R))_1(\sigma\cdot (L+L^R))_2
+(\sigma\cdot\tilde{L} +1)_1(\sigma\cdot \tilde{L} +1)_2
\right. \n
&&\left. 
-(\sigma\cdot (L+L^R))_1(\sigma\cdot \tilde{L} +1)_2
-(\sigma\cdot \tilde{L} +1)_1(\sigma\cdot (L+L^R))_2
\right] \ .
\label{chiraloperator12}
\eeqa
Acting it on the above states,
we obtain
\beq
\Gamma |1 \rangle
=  |2 \rangle \ ,  \ \
\Gamma |2 \rangle
=  |1 \rangle \ ,  
\eeq
where we used
$\frac{a^2}{4} \left[ (C_J)^2 
+\left( J+\frac{1}{2} \right)^2  \right]=1$.
We thus have
\beq
\Gamma \frac{1}{\sqrt{2}}\bigl(|1 \rangle +|2 \rangle \bigr)
= + \frac{1}{\sqrt{2}}\bigl(|1 \rangle +|2 \rangle \bigr)
 \ ,  \ \
\Gamma \frac{1}{\sqrt{2}}\bigl(|1 \rangle -|2 \rangle \bigr)
= -\frac{1}{\sqrt{2}}\bigl(|1 \rangle -|2 \rangle \bigr)  \ .
\eeq
The zero modes in this sector have both chiralities
and do not contribute to the index.

We next consider the case (b),
where two types of states $|J_1, + ;J_2, \pm \rangle$
exist.
Acting the GW Dirac operator (\ref{eqn:DGWfree})
on these states, and diagonalizing $D_{\rm GW}$,
we obtain the eigenstates as 
\beqa
&& D_{\rm GW} \
\frac{1}{\sqrt{2}}
\Bigl( | J_1, +, J_2, + \rangle \pm | J_1, +, J_2, - \rangle \Bigr)
\nonumber \\
&&=\pm \frac{a}{2} \left(J_1 + \frac{1}{2} \right) C_{J_2} \
\frac{1}{\sqrt{2}}
\Bigl( | J_1, +, J_2, + \rangle \pm  | J_1, +, J_2, - \rangle \Bigr) \ .
\eeqa
There is not a zero mode in this case.

We finally consider the case (c),
where only the states
 $| J_1, +; J_2, + \rangle$ exist.
Acting the GW Dirac operator (\ref{eqn:DGWfree})
and the chirality operator (\ref{chiraloperator12})
on these states,
we obtain
\beqa
D_{\rm GW} | J_1, +; J_2, + \rangle &=&0 \ , 
\label{eqn:zeromodeTtoT} \\
\Gamma | J_1, +; J_2, + \rangle 
&=& +| J_1, +; J_2, + \rangle \ .
\label{eqn:GG+TtoT}
\eeqa
They give chiral zero modes and contribute to the index.
Recalling that the state $| J_X, + \rangle$
has the $(2J_X+1)$-folded degeneracy,
the degeneracy of the chiral zero modes 
is $(2J_1+1)(2J_2+1)= 4n^2$.
This agrees with 
the first term in (\ref{ComLimTopCharResult}).

Now we have obtained the whole spectrum 
of the Dirac operator and checked that
the chiral zero modes are indeed given
by the states that we discussed in section 
\ref{sec:zero-mode_free}.

\section{Commutative limit in the modified formulation}
\label{sec:one_time_normalization}
\setcounter{equation}{0}

In this appendix, we consider the modified formulation
given by (\ref{defhatgamma_onetime}),
and calculate the commutative limit of the
Dirac operator and the topological charge.

By substituting (\ref{defH_albe}) into
(\ref{defhatgamma_onetime}),
and expanding it in $\beta$ and hence in the gauge fields, 
we obtain 
\beqa
&&\hat\Gamma'
\ = \ \al_1 \al_2 \nonumber \\
&&+ \Bigl( 
 \frac{1}{4} \{  \al_2, \ \be_1 - \al_1 \be_1 \al_1 \} 
+ ( 1 \leftrightarrow 2 ) \Bigr)  
\nonumber \\
&&+\frac{1}{32} \Biggl(
7\{ \be_1, \ \be_2 \} -5\al_1 \al_2\{ \be_1, \ \be_2 \}\al_1 \al_2
+3\{ \al_1 \be_1 \al_1, \ \al_2 \be_2 \al_2 \} \nonumber \\
&& -\Bigl(  
\{ \be_1, \ \al_2 \be_2 \al_2  \} 
+\al_1 \{ \be_1 , \ \be_2  \} \al_1
-3\al_1 \{ \be_1 , \ \al_2 \be_2 \al_2 \}\al_1
\nonumber \\
&& 
+ \al_2 \be_1 \al_2 \be_2 + \be_1 \al_1 \be_2 \al_1 
+\al_1 \al_2\be_1 \al_1 \al_2 \be_2 
+\al_2 \be_1 \al_1 \al_2 \be_2 \al_1
 \nonumber \\
&& 
+\al_1 \al_2\be_1 \al_2 \be_2 \al_1
+\al_2 \be_1 \al_1 \be_2 \al_1 \al_2
+\be_1 \al_1 \al_2\be_2 \al_1 \al_2
+ ( 1 \leftrightarrow 2 )  \Bigr) \nonumber \\
&& -\Bigl(
 \{ \al_2, \ \al_1 \be_1^2 +\be_1^2 \al_1+\be_1\al_1\be_1 \}
+\al_1 \be_1 \al_2 \be_1 +\be_1 \al_2 \be_1 \al_1 
+\be_1 \al_1 \al_2 \be_1 \nonumber \\
&& 
+\al_1 \al_2\be_1\al_2\be_1\al_2 
+\al_2 \be_1\al_2 \be_1\al_1\al_2
+\al_2 \be_1 \al_1\al_2 \be_1 \al_2
\nonumber \\
&&
-3(\{ \al_2, \ \al_1\be_1\al_1\be_1\al_1 \} 
+\al_1\be_1\al_1\al_2\be_1\al_1
+\al_1\al_2\be_1\al_1\al_2\be_1\al_1\al_2) 
 + ( 1 \leftrightarrow 2 )  \Bigr) \Biggr) \n
&&+{\cal O}(\be^3) \ . 
\label{eqn: hatGExpand_1nF}
\eeqa

The first and the second terms in (\ref{eqn: hatGExpand_1nF}),
which are 0-th and 1st order in $\beta$,
coincide with those of the original formulation,
(\ref{eqn:TrHatG_0}) and 
the 1st order terms 
in (\ref{eqn:TrHatG_1_1stChern1}),
at the operator level,
{\it i.e.}, before taking the trace.
Then, the commutative limit of the Dirac operator
$-a^{-1}(\Gamma-\hat\Gamma')$
becomes the same one as the original formulation,
(\ref{eqn:DGWcom}),
since
the commutative limit of the Dirac operator
is affected by $\hat\Gamma'$
only up to order $n^{-1}$.

We next consider the commutative limit of the
topological charge 
$\frac{1}{2}\CTr(\Gamma + \hat\Gamma')$,
which is affected by $\hat\Gamma'$
up to order $n^{-4}$.
While $\hat\Gamma'$ and $\hat\Gamma$ differ
at ${\cal O}(\be^2)$ at the operator level,
the trace of the difference becomes
\beqa
\CTr [\hat\Gamma'-\hat\Gamma ] &=&
\frac{1}{16} \CTr \Bigl[
[ \al_2, \ \be_1] \al_1 \al_2 [ \al_2, \ \be_1] 
-\al_1 [ \al_2, \ \be_1] \al_1\al_2 [ \al_2, \ \be_1] \al_1 
\Bigr] \n
&&
+ {\cal O}((\be_2)^2) + {\cal O}(\be_1\be_2) 
+{\cal O}(\be^3) \ ,
\label{eqn:CompareNew-Old}
\eeqa 
where we have written only the terms with $(\be_1)^2$. 
Since (\ref{eqn:CompareNew-Old}) vanishes in the commutative limit,
the commutative limit of the topological charge
$\frac{1}{2}\CTr(\Gamma + \hat\Gamma')$
becomes the same one as the original formulation,
(\ref{ComLimTopCharResult}). 

In the original formulation,
the commutative limit of
$\{\al_1, \ \zeta^{(2)}_2 \}$ and $\{\al_2, \ \zeta^{(2)}_1\}$
in (\ref{eqn:TrHatG_1_1stChern1}) 
gave the second order terms in the gauge field
in the 1st Chern character.
The commutative limit of
$\{\zeta^{(1)}_1, \ \zeta^{(1)}_2 \}$
in (\ref{eqn:TrHatG_2_2ndChern1}) gave 
the second order terms in the gauge field
in (\ref{eqn:2ndChern_1}),
which is a part of the 2nd Chern character . 
The commutative limit of
$\al_1\al_2([\al_1, \ \zeta_2^{(1)}]-[\al_2, \ \zeta_1^{(1)}])^2$
in (\ref{eqn:TrHatG_2_2ndChern2}) gave
the second order terms in the gauge field
in (\ref{eqn:2ndChern_2}).
However, in the modified formulation,
the corresponding terms are all mixed in the third term 
in (\ref{eqn: hatGExpand_1nF}),
and it is difficult to perform the same calculations
that we have done in the original formulation.
While the modified formulation is simpler in the 
definition since it has normalization procedure only one time,
calculations are easier in
the original formulation.


\end{document}